\documentclass[apl,preprint]{revtex4-1}
\usepackage{graphicx}
\usepackage{epstopdf}

\begin{document}
\title{Nonlinear motion and mechanical mixing in as-grown GaAs nanowires}
\author{F. R. Braakman$^{1}$}\thanks{Author to whom correspondence should be addressed.  Electronic mail: \texttt{floris.braakman@unibas.ch}.}
\author{D. Cadeddu$^{1}$}
\author{G. T\"ut\"unc\"uoglu$^{2}$}
\author{F. Matteini$ ^{2}$}
\author{D. R\"uffer$^{2}$}
\author{A. Fontcuberta i Morral$^{2}$}
\author{M. Poggio$^{1}$}
\affiliation{1: University of Basel, Klingelbergstrasse 82, 4056 Basel, Switzerland}
\affiliation{2: \'Ecole Polytechnique F\'ed\'erale de Lausanne, 1015 Lausanne, Switzerland}

\date{\today}

\begin{abstract}
We report nonlinear behavior in the motion of driven nanowire cantilevers. The nonlinearity can be described by the Duffing equation and is used to demonstrate mechanical mixing of two distinct excitation frequencies. Furthermore, we demonstrate that the nonlinearity can be used to amplify a signal at a frequency close to the mechanical resonance of the nanowire oscillator. Up to 26~dB of amplitude gain are demonstrated in this way. 
\end{abstract}
\maketitle
Due to their favorable geometry and potentially defect-free growth, nanowire cantilevers are promising as ultrasensitive force transducers for scanning probe microscopy~\cite{Nichol08,Nichol13,Feng07,Poggio13}. Additionally, their relatively high mechanical resonance frequencies decouple their motion to a large degree from external noise sources, and should permit improved sensitivity in mass-sensing and scanning probe applications. Furthermore, the wide choice of nanowire growth material and the possibility to grow nanowire heterostructures could allow access to different measurement modalities, such as sensing of local electric or magnetic fields. In recent experiments~\cite{Yeo14,Montinaro14}, coupling of optical transitions of a self-assembled quantum dot embedded in a nanowire to the motion of the nanowire through strain was demonstrated, opening the way to investigation of hybrid devices with nanowires as their main building blocks. Nanowire heterostructures are attractive as hybrid systems, as they can combine multiple functionalities in one integrated structure.

Conventionally, in scanning probe experiments oscillatory motion of the cantilever is driven with amplitudes small enough to remain in the linear dynamical regime. Due to a number of reasons~\cite{Villanueva13,Lifshitz08}, including the oscillator geometry, nonlinear damping~\cite{Eichler11,Zaitsev12}, the presence of external potentials, and nonlinear boundary conditions~\cite{Moon83,Tabaddor00}, this linear dynamic range is often quite limited in nanoscale oscillators~\cite{Antonio12,Postma05,Husain03}.
The nonlinear dynamics occurring when this range is exceeded complicate the analysis of sensing experiments and are therefore generally avoided or compensated for~\cite{Nichol09}. However, nonlinearities in general can also give rise to a host of useful effects, such as signal amplification~\cite{Almog06,Karabalin11}, noise squeezing~\cite{Almog07}, and frequency mixing~\cite{Erbe00}. The nonlinear dynamics of nanowire cantilevers can enable these effects at the nanoscale in mechanical form and have the potential to enhance the performance of cantilever-based sensors.

\begin{figure}[t]
\centering
\includegraphics{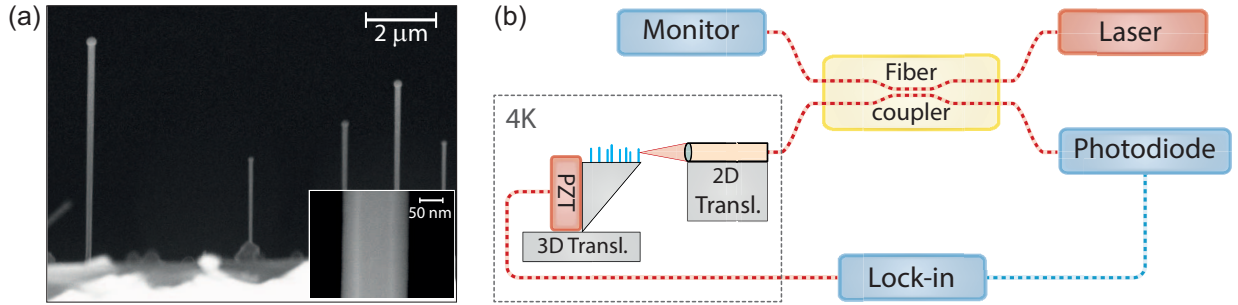}
\caption{(a) Scanning electron micrograph of a substrate containing multiple GaAs nanowires, taken at a different section of the same wafer that was used in the experiments. Inset: close-up of a single nanowire, showing a faceted structure due to its hexagonal cross-section. (b) Schematic diagram of the measurement setup.}
\label{fig:Figure1}
\end{figure}

In this Letter, we study the motion of several GaAs nanowires attached to their GaAs growth substrate (Fig.~\ref{fig:Figure1}(a)). 
We observe that, upon driving the periodic bending motion of a nanowire with sufficiently large amplitudes, it can no longer be described by a linear equation of motion. Instead, the nanowire follows the, qualitatively different, nonlinear dynamics of a Duffing oscillator~\cite{Nayfeh79}. A Duffing nonlinearity can give rise to complex motion of an oscillator, such as hysteresis, cascades of period-doubling, and chaotic motion~\cite{Barger95}. In the quantum regime, Duffing nonlinearities have recently been studied in the context of mechanical squeezing~\cite{Lue14}.
Furthermore, we find that when applying two driving frequencies, the nanowire motion in the nonlinear regime contains components at frequencies other than the two driving frequencies, as a result of mechanical mixing. 

The nanowires under investigation here are still attached to their GaAs growth substrate and are therefore singly clamped (see Fig.~~\ref{fig:Figure1}(a)). They have their fundamental mechanical resonances at frequencies of $f_0 = 1.25 - 1.35$~MHz (some nanowires show two closely spaced resonances, which we attribute to two transverse flexural modes that are non-degenerate due to a slight asymmetry of the nanowire cross-section) and exhibit quality factors of up to $37,000$ (at a temperature of 4.2~K and pressure below $10^{-6}$~mbar), as determined from ringdown measurements. 
The nanowires were grown on a 4 nm SiO$_\textrm{x}$ coated (111)B GaAs substrate by the catalyst-free Gallium-assisted method~\cite{Colombo08} in a DCA P600 solid source molecular beam epitaxy system. Growth has been done under a rotation of 7 rpm, with a growth rate of 0.5$\textrm{\AA}/$s and a substrate temperature of 630$^{\circ}$C. The nanowires mostly exhibit zinc-blende crystal structure, hence hexagonal cross-sections, with typical diameters of 100 nm and lengths up to 25 $\mu$m.

A schematic overview of the measurement setup is shown in Figure~\ref{fig:Figure1}(b). The displacement of a nanowire is measured via a fiber-based interferometric method~\cite{Hoegele08}. In this setup, the nanowire forms one reflecting interface of a low-finesse Fabry-P\'erot interferometer, while the cleaved surface of a single-mode fiber forms the other interface. The sample is mounted on a stack of positioning stages for three-axis translation control, allowing the nanowire of choice to be positioned in the focal plane of an objective placed in front of the single-mode fiber. A voltage-controlled piezoelectric transducer (PZT) attached to the sample holder is used to drive oscillatory bending motion of the nanowires along the optical axis of the interferometer. A fiber coupler is used to inject light from a laser with a wavelength of 1550~nm into the interferometer. This wavelength is chosen in order to avoid spurious heating of the GaAs nanowires through absorption. The coupler diverts $5\%$ of the laser power towards the nanowire, resulting in a maximum power incident on the nanowire of $\sim$5$\mu$W. The light reflected by the interferometer is collected by a photodiode with a dynamic range of 5~MHz. The oscillator of a lock-in amplifier actuates the PZT and the same lock-in amplifier demodulates the response of the photodiode. Sample and microscope are placed inside a vacuum can, which in turn is mounted inside a liquid helium bath cryostat.

\begin{figure}[t]
\centering
\includegraphics{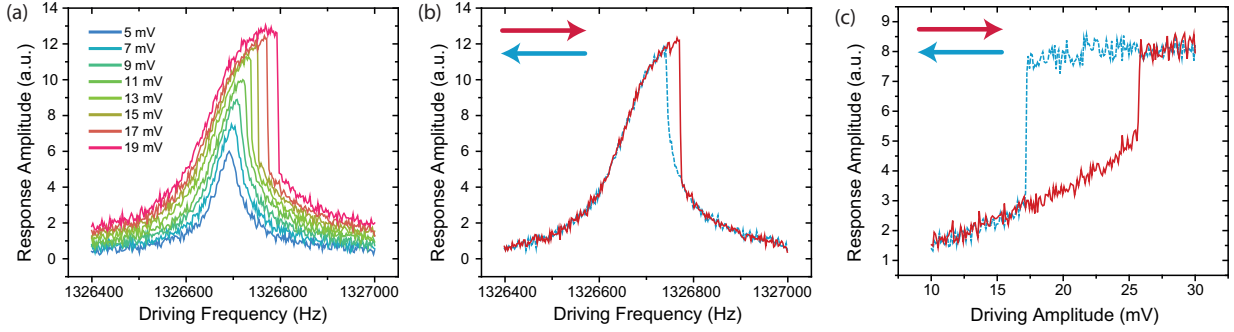}
\caption{(a). Response amplitude as a function of driving frequency, for various driving amplitudes. Note that the slight depression around the maximum response for the highest driving amplitude is caused by the limited linear range of the interferometer. (b) Response amplitude as a function of driving frequency (at a driving amplitude of 17~mV), for two sweep directions (as indicated by arrows). (c) Response amplitude as a function of driving amplitude (at a driving frequency of 1326770~Hz), for two sweep directions.}
\label{fig:Figure2}
\end{figure}

Figure~\ref{fig:Figure2}(a) shows the measured displacement of a nanowire for various driving amplitudes. As the driving amplitude is increased, the resonance becomes broader and assumes a characteristic shark-fin shape when entering the nonlinear regime, where the frequency associated with maximum displacement increases and moves away from the resonator eigenfrequency $f_0$. Such behavior is typical for a Duffing oscillator and can be described by the Duffing equation of motion:
\begin{equation}
\label{eq:motion}
\ddot{x}(t) + \mu \dot{x}(t) + {f_0}^2x(t)  + \alpha x^3(t) = F(t).
\end{equation}
Here $x$ is the displacement, $\mu$ the damping constant and $F(t)$ the time-dependent driving force, here taken to be sinusoidal. The coefficient $\alpha$ parametrizes the strength of the cubic nonlinearity. When $\alpha$ is positive, as it is in our case, the nonlinearity increases the effective spring constant with increasing driving amplitude, thus stiffening the motion. The observed lineshape at higher driving amplitudes is a consequence of Eq.~\ref{eq:motion} having two stable solutions within a certain frequency range. This bistability leads to the switching phenomena seen at the right flank of the response peak (Fig.~\ref{fig:Figure2}(a)). Which of the two solutions is realized, is determined by the initial conditions, and mechanical hysteresis can be observed when adiabatically sweeping the driving frequency or driving amplitude up and down (Figures~\ref{fig:Figure2}(b) and (c)). 
The strength of the nonlinearity $\alpha$ can be estimated from the shift of the frequency $f_{max}$ at which the maximum response amplitude occurs, using the relation~\cite{Nayfeh79}: $\alpha=\frac{32}{3}\pi^2f_0(f_{max}-f_0)/x^2$. 

Our interferometer becomes nonlinear for larger driving amplitudes (see Fig.~\ref{fig:Figure2}(a)), since then the displacement becomes comparable to the width of the interferometer fringes. We use this to infer~\cite{Dobosz98} a value for the displacement $x$ of $\sim$250~nm, for a driving amplitude of 19~mV (Fig.~\ref{fig:Figure2}(a)). We can then estimate $\alpha$ to be of order $10^{23}$m$^{-2}$s$^{-2}$ for this nanowire.

\begin{figure}[htb]
\centering
\includegraphics{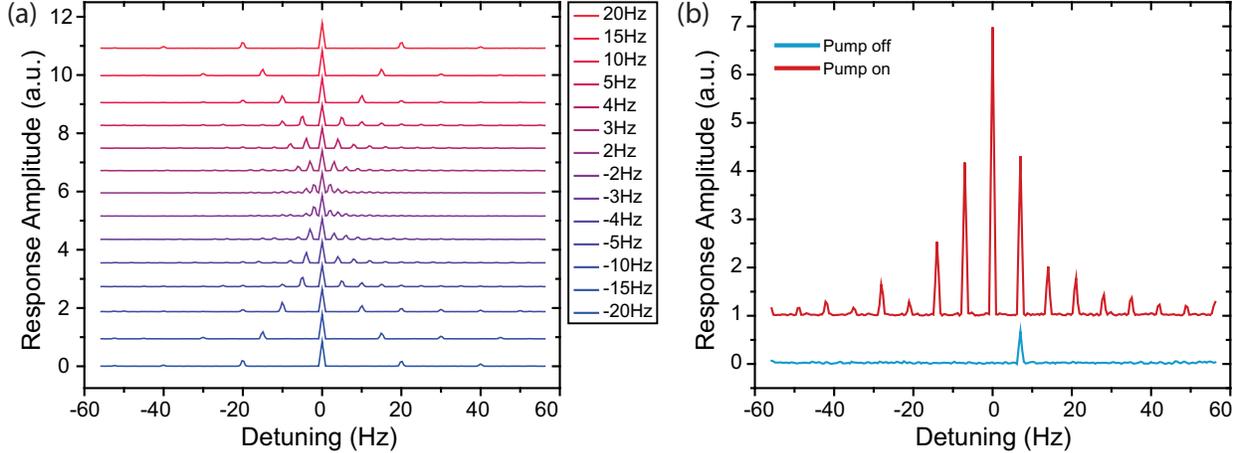}
\caption{(a) Spectral response amplitudes of the nanowire motion upon application of two driving frequencies, for various values of the detuning of the signal frequency from the pump frequency. The curves have been offset for clarity. The spectral reponse is given as a function of the detuning from the pump frequency, which is 1287890~Hz. The pump and signal amplitudes are 250~mV and 50~mV, respectively. The first mechanical mode of this nanowire has a resonant frequency of this nanowire at 1287780~Hz. (b) Spectral response with pump excitation (1287913~Hz, 250~mV) off (lower curve) and on (upper curve) for a signal detuned from the pump frequency by 7~Hz and with amplitude 35~mV. The curves have been offset for clarity.}
\label{fig:Figure3}
\end{figure}
Next, we show that the nonlinearity can be used to turn a nanowire into a mechanical mixer~\cite{Erbe00,Almog07}. Upon excitation with two driving frequencies, $F(t)=F_1~\textrm{cos}(f_1t+\phi_1)+F_2~\textrm{cos}(f_2t+\phi_2)$, the response shows sidebands additional to the motion at the driving frequencies, as shown in Figure~\ref{fig:Figure3}(a). We observe up to twelve such sidebands, spaced around the driving frequencies with splittings equal to the detuning between the two driving frequencies, $\Delta f=f_2 - f_1$. Note that these measurements were taken with a different nanowire than before, and that higher driving amplitudes were required to enter the nonlinear regime.

This response can be understood from Eq.~\ref{eq:motion} by taking the cubic term to be a perturbation to the driving force and solving the equation iteratively~\cite{Nayfeh79,Hutter10}. One then obtains new terms in the response at the intermodulation frequencies $f_1-n\Delta f$ and $f_2+n\Delta f$ (where $n$ is an integer) for each iteration. The amplitudes of these new intermodulation terms have coefficients proportional to $\sum\limits_{n} (f_0^2-f_1^2)^{-k}(f_0^2-f_2^2)^{-l}$, with $k$ and $l$ positive integers and $k+l=n$. Hence, intermodulation terms are smaller for driving frequencies that are more detuned from resonance. Since the mixing occurs due to the cubic term in Eq.~\ref{eq:motion}, for the intermodulation terms to be present, at least one of the driving amplitudes needs to be large enough to have an appreciable nonlinear response.

It is evident from Fig.~\ref{fig:Figure3} that the energy that is injected into the nanowire oscillator by the driving is distributed among the various intermodulation terms. This redistribution also occurs when one drive (signal) is much smaller than the other (pump), in which case amplification of the signal can take place~\cite{Almog06}. The signal here is formed by a driving voltage supplied to the PZT, but it could be any force driving the nanowire with a frequency close to the resonance. Fig.~\ref{fig:Figure3}(b) shows the spectral response of the nanowire motion with the signal drive always on, but with the pump excitation off in one case and on in the other. It is clear that amplification of the signal takes place when the pump excitation is switched on in the form of an increase in amplitude of the response at the signal frequency. Additionally, the appearance of the intermodulation terms, which is conditional on the presence of a signal, provides extra amplification. The total gain can be defined to be the ratio between the summed response amplitudes of all peaks present with pump drive, excluding the peak at the pump frequency itself, and the response amplitude with no pump drive. We observe a maximum gain of 26~dB.

In summary, we have observed and characterized nonlinear motion of as-grown GaAs nanowires. The nonlinearity is already observable for modest driving amplitudes. Furthermore, we have demonstrated that this nonlinearity allows for mechanical mixing of two excitations and amplification of a signal excitation through this mixing. This amplification could be utilized in several scanning probe techniques. For example, in the case where these nanowires act as mechanical force transducers, the observed gain of 26~dB could make force sensitivities of $\sim$100~zN/$\sqrt{\textrm{Hz}}$ in a narrow bandwidth feasible.
These results indicate that although nonlinear motion can be non-negligible for nanowires, the nonlinearity can also be turned into an advantage using simple measurement schemes. The nonlinearity could in addition lead to coupling of different flexural modes. Such nonlinear mode coupling could have several applications, including tuning the resonance frequency~\cite{Karabalin09} and quality factor~\cite{Venstra11} of one mode through driving of the other mode, and implementing quantum non-demolition measurements of mechanical excitation~\cite{Santamore04}.

\begin{acknowledgments}
We thank Pengfei Wang, Jiangfeng Du, and Sascha Martin for technical support. This work is supported by an ERC Grant (NWscan, Grant No. 334767), the Swiss Nanoscience Institute (Project P1207), QSIT, and the Sino Swiss Science and Technology Cooperation (Project IZLCZ2 1388904).
\end{acknowledgments}

\end{document}